\documentstyle[psfig]{l-aa}

\begin{document}

\newcommand{\beqa}{\begin{eqnarray*}}
\newcommand{\eeqa}{\end{eqnarray*}}
\newcommand{\beqan}{\begin{eqnarray}}
\newcommand{\eeqan}[1]{\label{#1}\end{eqnarray}}
\newcommand{\beq}{\begin{equation}}
\newcommand{\eeq}{\end{equation}}
\newcommand{\diff}{{\rm d}}
\newcommand{\drr}{\frac{\partial}{\partial r}}
\newcommand{\dtt}{\frac{\diff}{\diff t}}
\newcommand{\dr}[1]{\frac{\partial  #1}{\partial r}}
\newcommand{\dt}[1]{\frac{\partial  #1}{\partial t}}
\newcommand{\lp}{ \left(}
\newcommand{\rp}{ \right)}
\newcommand{\lc}{ \left[}
\newcommand{\rc}{ \right]}
\def\O{\Omega}

\bibliographystyle{plain}

\thesaurus{
02.08.1; 
08.09.3, 
08.18.1. 
}

\title{The Li dip : a probe of angular momentum transport in low mass stars}

\author{Suzanne Talon\inst{1} and Corinne Charbonnel\inst{2}}

\offprints{Suzanne Talon \\ Present address:
CERCA, 5160, boul. D\'ecarie, bureau 400,
Montr\'eal (Qu\'ebec) Canada H3X 2H9}

\institute{
D\'epartement d'Astrophysique Stellaire et Galactique, Observatoire de Paris,
Section de Meudon, 92195 Meudon, France 
\and Laboratoire d'Astrophysique de Toulouse, CNRS UMR5572, OMP,
14, Av. E.Belin, 31400 Toulouse, France \\
(Suzanne.Talon@obspm.fr, Corinne.Charbonnel@obs-mip.fr)}

\date{Received / Accepted }

\maketitle
\markboth{S. Talon and C. Charbonnel: 
The Li dip : a probe of angular momentum transport in low mass stars}{}

\author{Suzanne Talon\inst{1} and Corinne Charbonnel\inst{2}}

\offprints{Suzanne Talon}

\institute{
D\'epartement d'Astrophysique Stellaire et Galactique, Observatoire de Paris,
Section de Meudon, 92195 Meudon, France
\and Laboratoire d'Astrophysique de Toulouse, OMP,
14, Av. E.Belin, 31400 Toulouse, France \\
}{
}

\begin{abstract}

We use the measures of Li and rotational velocities in F Hyades stars 
to assess the role of the wind-driven meridian circulation 
and of shear turbulence in the transport of angular momentum in stars of 
different masses. 
Our models include both element segregation and rotation-induced mixing,
and we treat simultaneously the transport of matter and angular
momentum as described by Zahn (1992) and Maeder (1995).

We show that the hot side of the Li dip in the Hyades is well explained
within this framework, which was also successfully used to reproduce the C and N
anomalies in B type stars (Talon et al. 1997).
On the cool side of the dip, another mechanism must participate in the
transport of angular momentum; its 
efficiency is linked to the depth of the
surface convection zone.
That mechanism should also be responsible for the Sun's flat rotation profile.


\end{abstract}
\section{Towards a consistent description of rotation induced mixing}

During the last decade, special efforts have been devoted to 
improve the description of the mixing processes related to stellar 
rotation.
The most recent works (see for example Pinsonneault et al. 1989, Zahn 1992, 
Maeder 1995, Talon \& Zahn 1997) 
describe the evolution of the internal 
distribution of angular momentum in a self-consistent manner under the 
action of meridional circulation and of shear turbulence. 
The mixing of chemicals is then linked directly to the rotation profile, 
whereas previous studies made use merely of a parametric relation 
between the turbulent diffusivity and the rotational
velocity 
(cf. e.g. Schatzman et al. 1981, Zahn 1983).

Such a self-consistent treatment was applied successfully by 
Talon et al. (1997) in the study of a 9 M$_{\odot}$ star, modeling the
transport of angular momentum by the meridional circulation as a
truly advective process. 
The only assumption in this theory is that the turbulence sustained
by the shear is highly anisotropic and relies on two free parameters;
the first one describes the magnitude of the horizontal shears 
(cf. Zahn 1992) and
the second one, the erosion of the restoring force due to both the
thermal and the mean molecular weight stratifications (cf. Maeder 1995,
Talon \& Zahn 1997).
These authors reproduce the slight abundance anomalies measured in B 
stars by Gies \& Lambert (1992). 
They also show that the widening of the main sequence, which is generally
attributed to convective overshooting in massive stars, may be due to
the rotational mixing present in stars having a ``typical'' velocity
for the spectral type considered.

Concerning low-mass stars, 
it has been shown that the hydrodynamical models relying on meridional 
circulation and shear fail to reproduce the solar rotation profile 
given by the helioseismic observations
(Brown et al. 1989, Kosovichev et al. 1997): 
at the solar age, those models still have large $\Omega$ gradients 
which are not present in the Sun 
(see Chaboyer et al. 1995 and Matias \& Zahn 1997).
That conclusion has been reached independently by two different
groups, using different descriptions for the transport processes.
On one hand, 
the Yale group computed the evolution of angular momentum in low mass 
stars with a simplified description of the action of the 
meridional circulation which was considered as a diffusive process rather 
than as an advective process. The whole evolution of momemtum and
chemicals was then due to diffusion only, with a free parameter
that had to be calibrated to differentiate the transport of the
passive quantities with respect to that of vectorial ones. 
Pinsonneault et al. (1990) were then able to reproduce the surface Li 
abundances for low-mass cluster stars (with effective temperature lower
than 6500K). 
However, they obtained large rotation gradients within these
stars which are excluded by helioseismology (Chaboyer et al. 1995).
On the other hand, Matias \& Zahn (1997) performed a complete study
for the evolution of the Sun's angular momentum, where they took 
into account the advective nature of the meridional
circulation. They also concluded that meridional
circulation and shear turbulence are not efficient enough to enforce
the flat rotation profile measured by helioseismology.

These results indicate that another process participates in the 
transport of angular momentum in solar-type stars, while the so-called
wind-driven meridional circulation (Zahn 1992) is successful in more
massive stars. 
In order to study the transition between solar-type and more massive 
stars and to identify the mass range for which
the present description for the transport of angular momentum and
chemicals relying only on rotation fails, 
we propose to use the measures of lithium and rotational
velocities in galactic cluster stars. 

We first review the observations of lithium abundances and rotation in 
the Hyades main-sequence stars, and summarize the difficulties of the 
various models proposed so far to explain the Li dip in
F stars (\S 2). 
We recall the equations that describe the evolution of angular
momentum due to meridian circulation and shear turbulence as well as the 
associated transport of chemicals (\S 3).
We study the impact of rotational mixing on the lithium abundance in
galactic cluster F stars, and compare this to the observations.
Our models include both element segregation and rotation-induced mixing,
and we treat simultaneously the transport of matter and angular
momentum.
The internal rotation profile thus evolves completely self-consistently
under the action of meridional circulation as described by
Zahn (1992) (see also Matias et al. 1997), and of shear stresses
which take into account the weakening effect of the thermal
diffusivity, as was first shown by Townsend (1958) (\S 4). 
We show that the blue side of the lithium dip is well reproduced within
this framework, and that the process responsible for the shape of the solar
rotation profile should become efficient only for stars on the cool side
of the Li dip, where the external convection zone is thick enough. 
By achieving efficiently momentum transport, the global effect of 
this process would be to reduce the mixing due to the rotational
instabilities in stars with effective temperature lower than $\sim$
6500K. 
The most likely candidates for this transport process are 
the gravity waves generated by the external convection zone 
(Schatzman 1993; Zahn et al. 1997; Kumar \& Quataert 1997) and 
the large-scale magnetic field which could be present in the radiative 
interior (Charbonneau \& MacGregor 1993; Barnes et al. 1997). 
\section{Lithium and rotation in the Hyades F-stars}

Lithium (a fragile element which burns
at relatively low temperature in stellar interiors) has traditionally been
used as a very powerful tracer of particle transport processes.
Many Li abundance determinations are available for stars of different
spectral types in galactic clusters of various ages, together
with observed rotational velocities in some cases
(see for ex. Soderblom 1993, Balachandran 1995 and references therein).

Relying on [Li/Ca] observations in Hyades stars, 
Wallerstein et al. (1965) detected a drop-off in the 
lithium content of main sequence stars with a spectral index
around (B-V)=0.4.
It was clearly confirmed much later by Boesgaard \& Tripicco (1986) 
that Li is indeed depleted in Hyades F-stars in a range of 300 K in 
effective temperature centered around 6600 K (cf. Fig.~1 top). 
On the blue side of the so-called ``Boesgaard-dip", Li abundances drop 
sharply, while the rise on the red side is more gradual. 
Evidence of the same feature has been seen in all galactic clusters 
older than 10$^8$yr as well as in field stars (see Michaud \& Charbonneau 1991 
and Balachandran 1995 for a complete list of references).

The simplest explanation for this 
characteristic feature was proposed by Michaud (1986) who showed how chemical 
separation could shape the gap in the Hyades F stars. 
This model relied on well-known physics with two adjusted parameters: 
the mass loss rate needed to reduce the predicted over-abundances due to
radiative acceleration on the hot side of the plateau and 
the ratio of the mixing length to the pressure scale height. 
Three observational facts however contradict the pure microscopic 
diffusion hypothesis.
Firstly, the predicted width of the Li dip at the age of the Hyades is
narrower than observed (Richer \& Michaud 1993). 
Secondly, the carbon, oxygen and boron under-abundances expected in the
case of pure diffusion (Michaud 1986, Turcotte et al. 1997) 
failed to be found in the Hyades F stars (Boesgaard 1989, Friel \&
Boesgaard 1990, Garc\'\i a L\'opez et al. 1993) 
and in Li and Be deficient F field stars (Boesgaard et al. 1997).  
This indicates that a macroscopic process counteracts the effects of
element segregation in these stars.
Finally, in the pure diffusion model, Li settles and remains in a buffer 
zone below the convective envelope; it should then been 
dredged to the surface as soon as those stars leave the main sequence. 
Observations of lithium in M67's slightly evolved stars 
(Pilachowski et al. 1988; Balachandran 1995; 
Deliyannis et al. 1997) show however that the lithium depletion in stars 
formerly from the dip persists on the sub-giant branch.
This strongly favors explanations relying on nuclear destruction of lithium.

Schramm et al. (1990) proposed an explanation relying on mass loss.
The peeling of the outer layers of the stars could bring to the surface
the regions where lithium has been depleted by pure nuclear destruction. 
However, the existence of Hyades and field F stars which still have 
some lithium but where some beryllium has been also depleted argues 
against that mechanism (Stephens et al. 1997).

\begin{figure}[t]
\centerline{
\psfig{figure=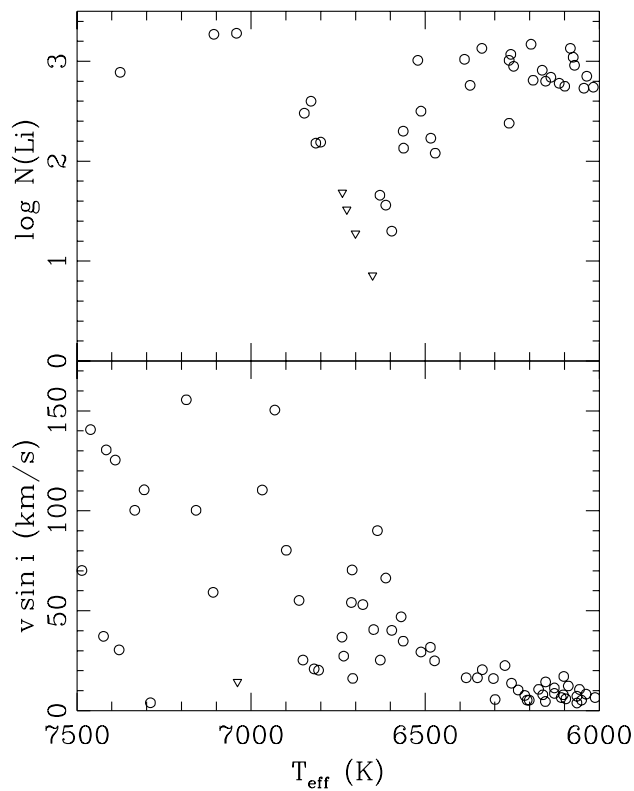,height=11cm}
}
\caption{({\it top}) Lithium versus effective temperature in F Hyades
stars (Boesgaard \& Tripicco 1986, Thorburn et al. 1993).
Triangles denote upper limits.
({\it bottom}) Projected rotational velocity versus effective
temperature in F Hyades stars.  Observational data are from Kraft (1965), 
Stauffer et al. (1987), and
Mermilliod (1992).}
\end{figure}

Following Press' suggestion (1981) that gravity waves may induce shear 
mixing, Garc\'\i a L\'opez \& Spruit (1991) studied the transport of 
lithium as a function of spectral type. 
They found that low degree waves may become efficient sources of shear 
mixing for stars of the Li dip, as their production model is mainly 
dependent on the convective flux. A suitable choice of the mixing
length then permits one to place the dip at the correct effective temperature
(since it influences the disappearance of the convective zone).
However, they needed to increase the efficiency of the wave generation
by a factor of 15 over their estimation to correctly reproduce the
dip. Furthermore, that model fails to reproduce the large abundance 
dispersion observed on the red side.

Boesgaard (1987) noticed that the Li dip in the Hyades coincides 
(in terms of effective temperature) with both a sharp drop in rotational 
velocities (see Fig.~1 bottom) and with the transition from high stellar 
activity to activity controlled by the stellar dynamo 
(Wolff et al. 1986). 
Rotation was then suggested to play a dominant role in the build up
of the Li dip.
Up to now, the different investigations of the possible connection 
between rotation and Li deficiencies in F stars have relied on highly 
simplified descriptions of the rotation-induced mixing processes. 
In the meridional circulation model of Tassoul \& Tassoul (1982) used by 
Charbonneau \& Michaud (1988), the feed-back effect due to angular 
momentum transport as well as the induced turbulence were ignored. 
Following Zahn (1992), Charbonnel et al. (1992, 1994) considered the 
interaction between meridional circulation and turbulence induced by 
rotation, but the transport of angular momentum was not treated 
self-consistently.

Here we go one step further by including in our models the most complete 
description currently available for rotation-induced mixing, and 
we compute simultaneously the transport of chemicals and the
transport of angular momentum due to wind-driven meridional circulation. 
Let us stress again that this description has been used to
successfully reproduce the slight over (under) abundances of C
(N) observed in B type stars (cf. Talon et al. 1997)
but failed
to explain the flat rotation profile observed in the Sun
(cf. Matias \& Zahn 1997). 
Another process which efficiently transports
angular momentum must be invoked in order to explain the helioseismic data.
Here, we use the observations of lithium and rotation in the Hyades in
order to get constraints on the onset of that process.
\section{The transport processes : Rotational mixing and microscopic
diffusion}

We calculate the destruction of lithium in F-type stars, assuming that
rotational mixing is the only source of transport for angular momentum. 
The evolution of the interior radial differential rotation is calculated
completely self-consistently, using the most complete description 
currently available for the following physical processes:
\begin{itemize}
\item
the advection of angular momentum by the meridional
flow driven by the thermal imbalance in a rotating star,
assuming that the rotation velocity is homogenized on isobars
by anisotropic shear turbulence, as described by Zahn (1992);
\item
the turbulent transport due to the
vertical shear present in differentially rotating bodies,
including the weakening effect of the thermal diffusivity on the
density stratification (for details see e.g. Talon \& Zahn 1997).
\end{itemize}

The complete equation for the transport of angular momentum is
then
\beq
\rho \dtt \lc r^2 {\O}\rc = 
\frac{1}{5 r^2} \drr \lc \rho r^4 {\O}
U \rc + \frac{1}{ r^2} \drr \lc \rho \nu_v r^4 \dr{\O} \rc
\label{ev_omega}
\eeq
where we use standard notations for the radius $r$ and
the density $\rho$ and where
$\nu_v$ is the vertical (turbulent) viscosity.
$U(r)$ is the vertical component of the meridian velocity and is
given by
\beq
U(r) = {L \over M g} \left( {P \over C_P \rho T} \right)
 {1 \over \nabla _{\rm ad} - \nabla} \lc E_{\O} + E_{\mu} \rc ,
\eeq
where $L$ is the luminosity, $M$ the mass, $g$ the gravity, $P$
the pressure, $C_P$ the specific heat at constant pressure and $T$
the temperature.
$E_{\O}$ and $E_{\mu}$ depend respectively on the rotation profile
and on the mean molecular weight gradients (for the complete
expression, see Zahn 1992).
The expression of the turbulent viscosity is
\beq
\nu _v = \frac{2}{5} K \lp \frac{r}{N} \frac{\partial
\Omega}{\partial
r} \rp ^2,
\label{dift}
\eeq
where $N$ is the Brunt-V\"ais\"al\"a frequency and $K$ is 
the thermal diffusivity.
The coefficient $\frac{2}{5}$ used here is the one that was found
by Maeder (1995) when he rederived the criterion for shear
instabilities assuming spherical geometry for the turbulent
eddies. As was discussed by Talon \& Zahn (1997), even though this is
somewhat of an arbitrary choice, the exact value
shouldn't differ much.
In this study, we will use the value $\frac{2}{5}$ and not
consider it as a free parameter.

Microscopic diffusion of lithium, helium and metals,
including gravitational and thermal settling, is taken into account 
(see Appendix for a description of the corresponding input physics).

Modeling the combination of the advective transport by the
circulation and the strong horizontal diffusion $D_h$ present in stratified 
media by an effective diffusivity $D_{\rm eff}$ (cf. Chaboyer \& Zahn 1992):
\beq
D_{\rm eff} = {|r U(r)|^2 \over 30 \, D_h} \; ,
\label{deff}
       \eeq
the evolution of a chemical concentration $c_i$ is given by:
\beqan
\rho {\partial c_i \over \partial t}  =&& \hspace{-0.2cm}
\dot{c_i} +
{1 \over r^2} {\partial \over \partial r} \left[r^2
 \rho \, U_{\rm diff} \, c_i  \right]  \nonumber \\
&&+ {1 \over r^2} {\partial \over \partial r}
\left[ r^2 \rho \left(D_{\rm eff} + D_{\rm turb} \right)
{\partial c_i \over \partial r} \right]  ,
\eeqan{ev_chem}
where $\dot{c_i}$ is the nuclear production/destruction rate and
$U_{\rm diff}$
is the microscopic diffusion; we assume $D_{\rm turb}=\nu_v$.
The weakest point is this model is the magnitude of the horizontal
diffusion coefficient. Here, we will use a parametric relation which links
that coefficient to the advection of momentum:
\beq
D_h = \frac{rU}{C_h} \lc \frac{1}{3} \frac{{\rm d} \ln \rho r^2 U}
{{\rm d} \ln r} - \frac{1}{2} \frac{{\rm d} \ln r^2 \O}{{\rm d} \ln r} \rc,
\eeq
where $C_h$ is an unknown parameter of order unity (see Zahn 1992 for more
details).
\section{The efficiency of rotational mixing in F-stars}

\begin{figure}
\centerline{
\psfig{figure=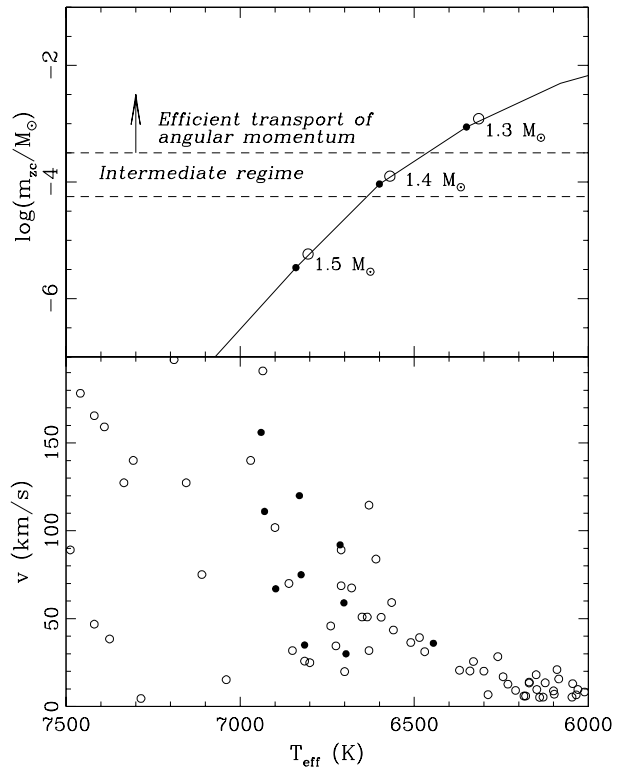,height=11cm}
}
\caption{({\it top}) The size of the convective envelope ($m_{zc}$)
at the age of the Hyades is plotted as a function of the stellar effective
temperature. The filled dots represent standard models (i.e., non rotating);  
the open dots correspond to rapidly rotating models (150 km/s, not spun
down at the age of the Hyades). Fast rotators have lower effective
temperatures and deeper convective envelopes than the standard models. 
({\it bottom}) Rotational velocities of our models at the age of the Hyades 
(filled dots; see Table 1). Also shown are the measured
projected velocities times 4/$\pi$ (open dots).
\label{basezc}
}
\end{figure}

\begin{figure}
\centerline{
\psfig{figure=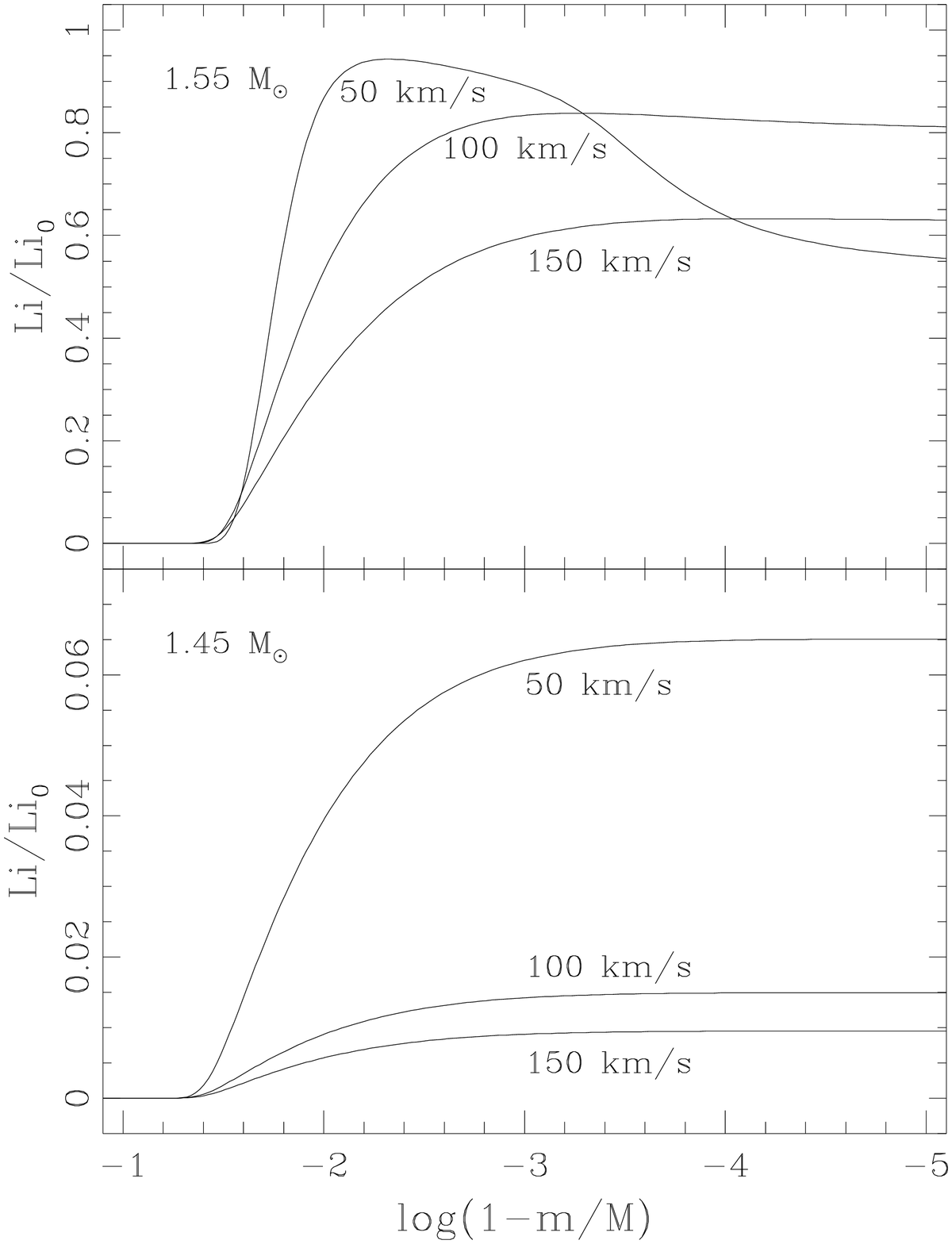,height=11cm}
}
\caption{Interior Li profile at the age of the Hyades for two different
masses (1.55 and 1.45 M$_{\odot}$) and three different initial
velocities (50, 100 and 150 km/s). 
The 1.55 M$_{\odot}$ model is not spun down and conserves its initial
angular momentum; only the slower model feels the effect of microscopic
diffusion, while rotation induced mixing dominates below the convection
zone of the fastest models. 
The 1.45 M$_{\odot}$ model is strongly spun down and Li destruction 
is large for all values of the initial velocity.
\label{profilli}
}
\end{figure}

\subsection{Constraints from the observed surface rotation in
galactic cluster F stars}

As can be seen in Fig.~1, the stars of the Li dip are peculiar as far as
their rotational history is concerned.
From the observational data, one may conclude that the physical processes 
responsible for the surface velocity are different or operate with different 
time scales when one goes to lower effective temperature:
stars hotter than $\sim$ 6900 K still have their initial
velocities (when compared to the velocity distribution observed in
younger clusters)
while stars cooler than $\sim$ 6400 K have been very efficiently spun
down at the age of the Hyades (700 Myr).

\begin{table*}
\caption{Characteristics of the stellar models, rotation velocities and lithium 
and beryllium depletion factors at the ages of the considered galactic open 
clusters. The models are computed with [Fe/H] of the Hyades (+0.12)}
\begin{center}
\begin{tabular}{ccccccccc}
\hline
\multicolumn{1}{c}{$M_*$}
& \multicolumn{1}{c}{t}
& \multicolumn{1}{c}{$T_{\rm eff}$}
& \multicolumn{1}{c}{log$\left({{L_*}/{L_{\odot}}}\right)$}
& \multicolumn{1}{c}{log$\left({m_{zc}/{M_*}}\right)$}
& \multicolumn{1}{c}{$T_{zc}$}
& \multicolumn{1}{c}{v}
& \multicolumn{1}{c}{Li/Li$_0$}
& \multicolumn{1}{c}{Be/Be$_0$}\\
(M$_{\odot}$) & (Myr) & (K) & & & (10$^6$ K) & (km s$^{-1}$) & & \\
\hline
1.55 & 50 & 7045 & 0.71 & -7.5 & 0.084 & 70 & 0.57 & 0.62 \\
     &100 & 7040 & 0.71 & -7.5 & 0.079 & 70 &0.30 & 0.35 \\
     &300 & 6990 & 0.72 & -7.5 & 0.081 & 70 &0.31 & 0.36 \\
     &450 & 6950 & 0.73 & -7.3 & 0.087 & 70 &0.44 & 0.48 \\
     &700 & 6900 & 0.75 & -6.8 & 0.12  & 65 &0.54 & 0.58 \vspace{0.1cm}\\
     & 50 & 7045 & 0.71 & -7.5 & 0.084 & 120 & 0.53 & 0.57\\
     &100 & 7035 & 0.71 & -7.5 & 0.082 & 120 & 0.45 & 0.49\\
     &300 & 7020 & 0.73 & -7.1 & 0.099 & 115 & 0.78 & 0.81\\
     &450 & 6985 & 0.74 & -7.0 & 0.11  & 110 & 0.82 & 0.85\\
     &700 & 6925 & 0.75 & -6.6 & 0.13  & 110 & 0.79 & 0.85 \vspace{0.1cm}\\
     & 50 & 7045 & 0.71 & -7.5 & 0.083 & 165 & 0.54 & 0.57\\
     &100 & 7050 & 0.71 & -7.3 & 0.093 & 165 & 0.80 & 0.83 \\
     &300 & 7025 & 0.73 & -7.1 & 0.11  & 160 & 0.89 & 0.90 \\
     &450 & 7000 & 0.74 & -7.0 & 0.11  & 155 & 0.82 & 0.90 \\
     &700 & 6940 & 0.75 & -6.7 & 0.12  & 155 & 0.60 & 0.83 \vspace{0.1cm}\\
1.5  & 50 & 6900 & 0.64 & -5.8 & 0.23 & 45 & 0.96 & 0.99 \\
     &100 & 6900 & 0.65 & -5.8 & 0.23 & 45 & 0.85 & 0.98 \\
     &300 & 6880 & 0.66 & -5.8 & 0.22 & 40 & 0.43 & 0.82 \\
     &450 & 6860 & 0.67 & -5.7 & 0.23 & 40 & 0.26 & 0.66 \\
     &700 & 6815 & 0.69 & -5.6 & 0.24 & 35 & 0.12 & 0.46 \vspace{0.1cm}\\
     & 50 & 6900 & 0.64 & -5.8 & 0.23 & 95 & 0.89 & 0.99 \\
     &100 & 6900 & 0.65 & -5.8 & 0.23 & 90 & 0.74 & 0.96 \\
     &300 & 6880 & 0.66 & -5.8 & 0.22 & 85 & 0.35 & 0.74 \\
     &450 & 6865 & 0.67 & -5.8 & 0.22 & 80 & 0.19 & 0.56 \\
     &700 & 6825 & 0.69 & -5.6 & 0.23 & 75 & 0.076 & 0.35 \vspace{0.1cm}\\
     & 50 & 6900 & 0.64 & -5.8 & 0.23 &145 & 0.85 & 0.98 \\
     &100 & 6900 & 0.65 & -5.8 & 0.23 &140 & 0.69 & 0.95 \\
     &300 & 6885 & 0.66 & -5.8 & 0.22 &130 & 0.30 & 0.69 \\
     &450 & 6865 & 0.67 & -5.8 & 0.22 &125 & 0.16 & 0.50 \\
     &700 & 6830 & 0.69 & -5.7 & 0.23 &120 & 0.060 & 0.30 \vspace{0.1cm}\\
1.45 & 50 & 6745 & 0.58 & -4.9 & 0.36 & 45 & 0.96 & 0.99\\
     &100 & 6745 & 0.58 & -4.9 & 0.36 & 45 & 0.83 & 0.98 \\
     &300 & 6735 & 0.59 & -4.9 & 0.36 & 35 & 0.35 & 0.76 \\
     &450 & 6725 & 0.60 & -4.9 & 0.36 & 35 & 0.17 & 0.55\\
     &700 & 6695 & 0.62 & -4.8 & 0.36 & 30 & 0.058 & 0.32\vspace{0.1cm}\\
     & 50 & 6745 & 0.58 & -4.9 & 0.36 & 90 & 0.89 & 0.99 \\
     &100 & 6745 & 0.58 & -4.8 & 0.36 & 85 & 0.68 & 0.95 \\
     &300 & 6740 & 0.59 & -4.8 & 0.36 & 70 & 0.18 & 0.55 \\
     &450 & 6725 & 0.60 & -4.8 & 0.35 & 65 & 0.063 & 0.32 \\
     &700 & 6700 & 0.62 & -4.8 & 0.36 & 60 & 0.012 & 0.14 \vspace{0.1cm}\\
     & 50 & 6745 & 0.58 & -4.9 & 0.36 &140 & 0.84 & 0.98 \\
     &100 & 6745 & 0.58 & -4.9 & 0.36 &130 & 0.60 & 0.91 \\
     &300 & 6740 & 0.59 & -4.9 & 0.35 &110 & 0.12 & 0.48 \\
     &450 & 6730 & 0.60 & -4.9 & 0.35 &105 & 0.040 & 0.25 \\
     &700 & 6715 & 0.62 & -4.9 & 0.34 & 90 & 0.008 & 0.10 \vspace{0.1cm}\\
1.35 & 50 & 6470 & 0.43 & -3.5 & 0.74 & 80 & 0.98 & 0.99 \\
     &100 & 6470 & 0.44 & -3.6 & 0.72 & 70 & 0.96 & 0.97 \\
     &300 & 6465 & 0.45 & -3.6 & 0.70 & 50 & 0.89 & 0.91 \\
     &450 & 6460 & 0.46 & -3.6& 0.68  & 40 & 0.83 & 0.86 \\
     &700 & 6445 & 0.47 & -3.6 & 0.67 & 35 & 0.74 & 0.77 \\
\hline
\end{tabular}
\end{center}
\end{table*}

This behavior is linked to the variation of the thickness of the
external convection zone (Fig.~\ref{basezc} top). 
Indeed, the hottest stars have only a very shallow surface convection zone
which is not an efficient site for magnetic generation via a dynamo process.
The coolest stars have a deeper surface convection zone, thus sustaining
a strong magnetic field which spins down the outer layers efficiently.
The rapid diminution with increasing effective temperature in the efficiency 
of magnetic braking observed for stars of the Li dip is a clear signature of 
the rapid decrease in mass of the envelope convection zone in stars of the 
corresponding effective temperature. 
Let us note that the diminution of the moment of inertia of the
convective envelope as the effective temperature increases
implies an even more drastic change in the magnitude of the magnetic
torque than the variation of the surface velocities indicate.

\begin{figure}
\centerline{
\psfig{figure=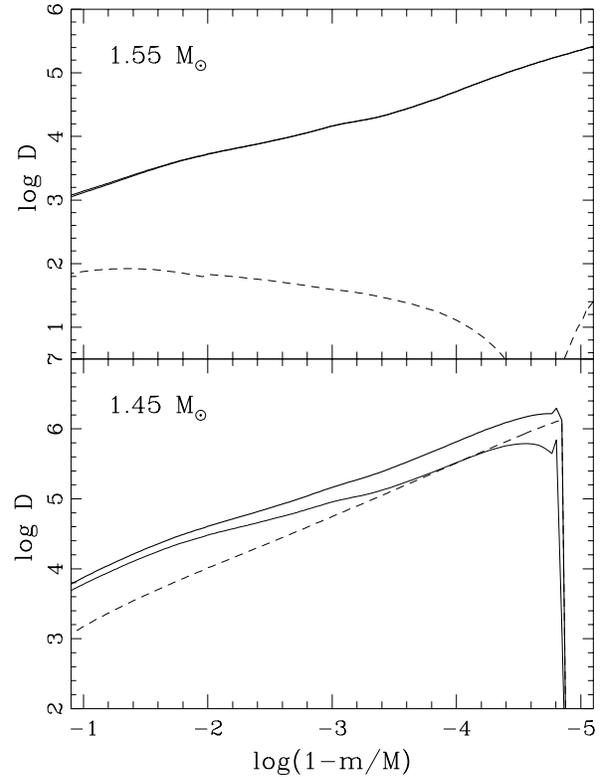,height=11cm}
}
\caption{Interior profile of the diffusion coefficient at the age of the
Hyades and for an initial velocity of 100 km/s. 
The wide full line represents the total diffusion coefficient,
the thin full line, the turbulent diffusion coefficient (cf. Eq.
\ref{dift}) and the dashed line, the effective diffusion coefficient
(cf. Eq. \ref{deff}).
The 1.55 M$_{\odot}$ model is in the asymptotic regime, and it is
turbulence which dominates the transport of chemicals whereas 
the transport of angular momentum by turbulence and meridional circulation
is of equal magnitude and opposite signs.
For the 1.45 M$_{\odot}$ model, while the transport of angular momentum
is done mainly by the circulation, the transport of chemical is dominated
by the former only close to the surface.
\label{profilcd}
}
\end{figure}

\begin{figure}
\centerline{
\psfig{figure=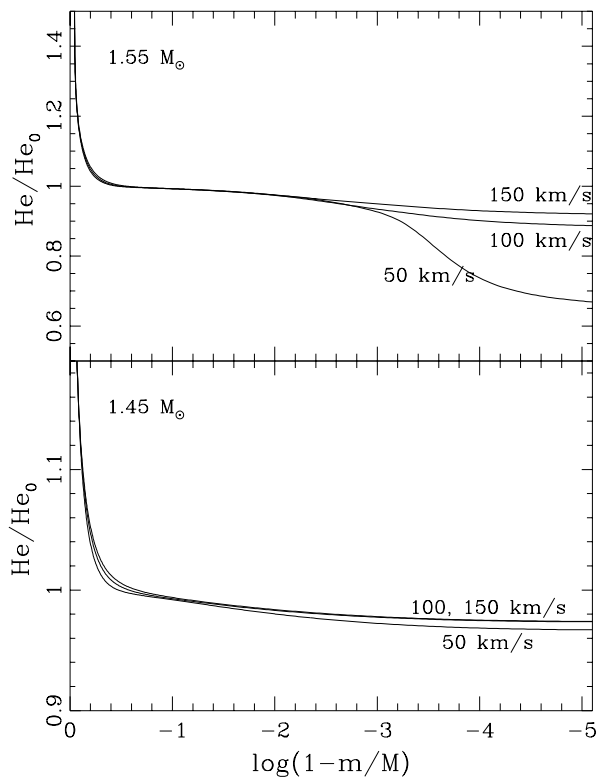,height=11cm}
}
\caption{Interior He profiles at the age of the Hyades, for the same
models and rotation velocities as in Figure~3. 
\label{profilhe}
}
\end{figure}

\begin{figure}
\centerline{
\psfig{figure=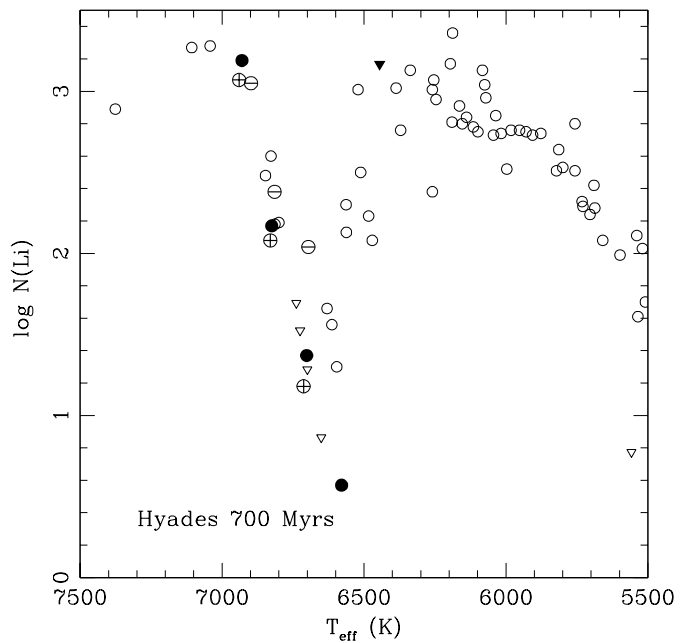,height=9cm}
}
\caption{
Comparison of the models (filled dots and triangle, pluses and minuses)
with the observations in the Hyades (open dots and triangles).
The model with $T_{\rm eff} \simeq$ 6900 K conserves its global angular
momentum during its main sequence life while for models with
6500 $<T_{\rm eff}({\rm K})<$ 6900, angular momentum is lost at the surface.
Interior redistribution of momentum takes place only through rotational
mixing; the coolest model ($T_{\rm eff} \simeq$ 6450 K) is computed
assuming solid body rotation.
The filled dots correspond to numerical calculations performed with an
initial velocity of 100 km/s, the pluses to an initial velocity of 150 km/s,  
and the minuses to initial velocity of 50 km/s. The corresponding
rotational velocities at the age of the Hyades are shown in Fig.~2 (see
also Table 1).
Numerical calculations are performed using a value of [Fe/H]=+0.12,
i.e., that of the Hyades.
\label{livsteff}
}
\end{figure}

We calculate Li destruction in models of different stellar masses within the
theoretical framework described in \S 3. 
We use the statistical study of rotation velocities in the Hyades
performed by Gaig\'e (1993) in order to estimate the spin down
associated to stars of different masses:
we take an initial velocity of 100 km/s that corresponds to the mean velocity 
of hot stars ($T_{\rm eff}>7000$ K) and that is consistent
with velocities of cooler stars measured in younger clusters.
The resulting velocity at the age of the Hyades (Fig.~\ref{basezc}
bottom) corresponds to the average value for stars of a given effective
temperature. 
We also calculate the dispersion expected in the Li abundances from
different rotational histories using the $\pm 1\sigma$ values from
Gaig\'e's study. The corresponding initial velocities are
50 and 150 km/s.
The main characteristics of the stellar models, together with the rotation 
velocities and the lithium and beryllium depletion factors at different ages 
(corresponding to the ages of the clusters shown in Figs.~6 and 7) 
are given in Table 1. 

\subsection{The depletion on the blue side of the Li dip}

\noindent{\bf {Stars hotter than $\sim$6900 K}}
\vspace{0.1cm}

\noindent According to the observations, stars hotter than $\sim$6900 K 
are not slowed down by a magnetic torque.
Since equation (\ref{ev_omega}) admits a stationary solution, those stars
soon reach a regime with no net angular momentum flux, in which
meridional circulation and shear turbulence counterbalance each
other.
The weak mixing resulting from these processes is just sufficient to
counteract the effect of microscopic diffusion, except in the slowest
rotators in which its signature is visible
(cf. Fig.~\ref{profilli} top and Fig.~\ref{profilcd} top for the shape
of the diffusion coefficients in the 1.55M$_{\odot}$ model).
In our calculations of microscopic diffusion, 
only gravitational settling was included
whereas in these stars, lithium may be supported by
radiative acceleration (Michaud 1986).
In improved radiative force calculations, Richer et al. (1997)
showed however that the actual force on Li critically depends on the 
helium and metals abundance; in the dip region, gravitational settling of 
helium is needed for lithium to be supported against gravity. 
As can be seen in Fig.~\ref{profilhe}, the mixing strongly slows down 
the diffusion of helium in the external layers of the star. 
This may lead to a significantly reduction of the radiative 
acceleration on Li on the hot side compared to previous estimations. 
In any case, the main effect is that turbulence considerably reduces 
microscopic diffusion, and leads only to small variations in the surface 
abundances.
\vspace{0.1cm}

\begin{figure*}
\centerline{
\psfig{figure=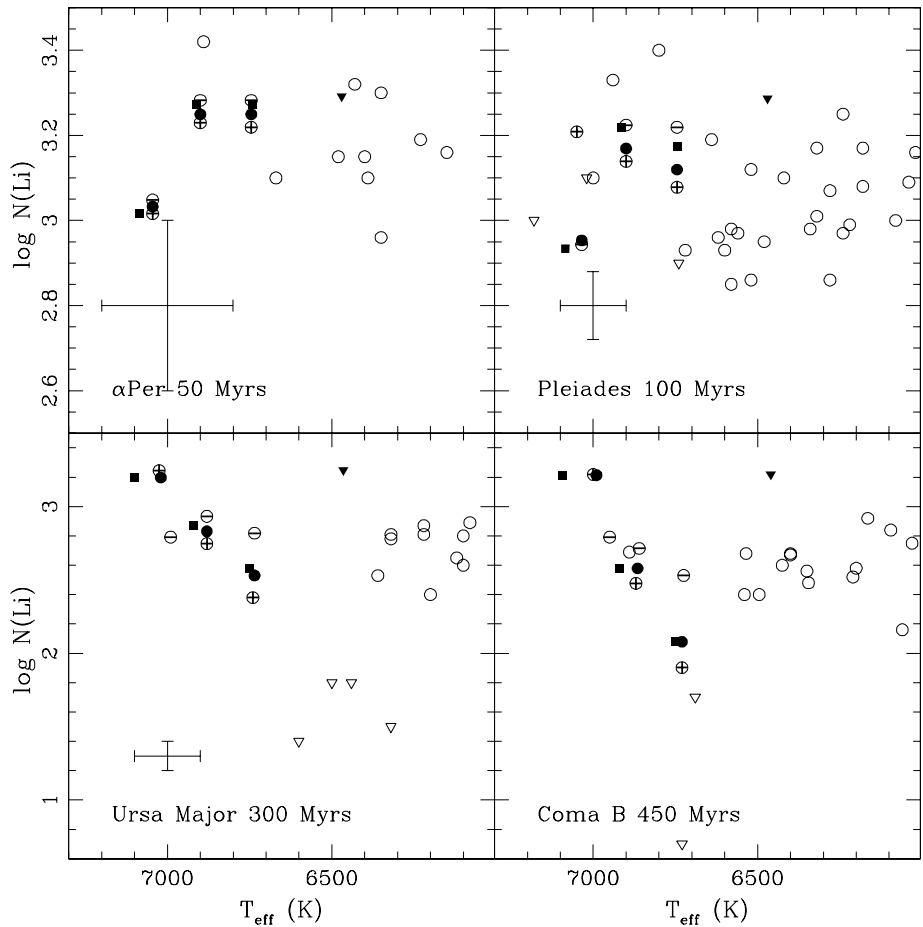,height=13.5cm}
}
\caption{Shape of the Li dip at the age of different
galactic clusters. 
Symbols are as in Fig.~\ref{livsteff}.
Filled squares denote calculations made for a metallicity of [Fe/H]=$-$0.15
and an initial velocity of 100 km/s.
Typical observational error bars are shown when available.
\label{amas}
}
\end{figure*}

\noindent{\bf {Stars with 6600$<T_{\rm eff}({\rm K})<$6900}}
\vspace{0.1cm}

\noindent 
In this effective temperature range,
a weak magnetic torque slows down the outer layers of the stars. 
As the magnetic torque increases with diminishing temperatures,
meridional circulation has much more angular momentum
to transport to the surface, leading to a larger destruction of Li
(cf. Fig.~\ref{profilli} bottom and Fig.~\ref{profilcd} bottom
for the shape of the 
diffusion coefficients in the 1.45M$_{\odot}$ model).
Again, Li is transported by meridional circulation, shear turbulence
and microscopic diffusion. 
Let us note however that here, the gap is not shaped by element
segregation, as in Michaud (1986) but by the properties of braking 
and angular momentum transport.
Furthermore, since here Li is destroyed and not barely hidden below the
convection zone, no dredge-up of this element is expected due to
the deepening of the external convection zone when the stars will leave
the main sequence. This is in agreement with the absence of Li detection
in Balachandran's (1995) sub-giants. 

The predictions for the lithium abundance at different ages are given in 
Figs.~6 and 7, and compared to observations in galactic clusters. We
also show predictions for models with [Fe/H]=-0.15 in Fig.~7, in order to
take into account the metallicity differences between the various
clusters. 
At the age of the Hyades, rotational mixing described in \S 3
perfectly explains the shape of the blue side of the Li dip, 
as well as the observed dispersion. 
This clearly indicates that, in this effective temperature range, the 
process which participates to the transport of angular momentum in the
Sun is not yet efficient. 

We computed the effect of rotational mixing on beryllium, which burns at
a slightly higher temperature than lithium ($\sim$ 3.5 $\cdot$ 10$^6$K 
instead of $\sim$ 2.5 $\cdot$ 10$^6$K). 
At the age of the Hyades, the surface beryllium abundance has diminished by 
as much as a factor 5 in the center of the dip. 
This is in agreement with the observations by Boesgaard \& Budge (1989) in
the Hyades and by Stephens et al.(1997) in field stars, as can be seen
in Figure~8. However, the comparison with Stephens' data is marginal 
because of the inhomogeneity in metallicity and evolutionary status of 
their sample. 
New observational data, with modern detectors, would be necessary for
beryllium in more stars of the Li dip. 

\subsection{The red side of the Li dip}

\noindent {\bf {Stars with 6400$<T_{\rm eff}({\rm K})<$6600}}
\vspace{0.1cm}

\noindent On the red side of the dip, the magnetic torque strengthens 
as the convective zone grows (Fig.~\ref{basezc}). 
If we assume there that all the momentum transport is assured by the 
wind-driven meridional circulation, and if we keep the same parameters that
explain the Li abundances on the blue side of the dip as well as the
chemical anomalies in more massive stars, we obtain too much lithium
burning compared to the observations in this cool effective temperature
domain (cf. Fig.~\ref{livsteff}, model with $T_{\rm eff} \simeq$ 6550 K).
Even though a different calibration of the free parameters may lead to
the observed lithium abundances, 
it would not change the internal rotation profile which is
known to be inconsistent in the solar case. 

We rather propose that the red side of the dip corresponds to a transition 
region where some other physics for angular momentum transport, which is 
known to be present in the Sun, starts to become efficient. 
In that case, the magnitude of both the meridional circulation and shear 
turbulence is reduced, as well as the Li depletion by rotation-induced mixing. 
We suggest that this increase of efficiency is linked to the growth of the 
surface convection zone (see Fig.~\ref{basezc} top).
This efficient transport mechanism for angular momentum could be due to
gravity waves or to a magnetic field in the radiative interior
(see references given in the first part).

A complete description of the efficient mechanism for the transport of angular 
momentum is required in order to calculate self-consistently the Li destruction 
in this region, and has not been attempted here.
\vspace{0.1cm}

\begin{figure}
\centerline{
\psfig{figure=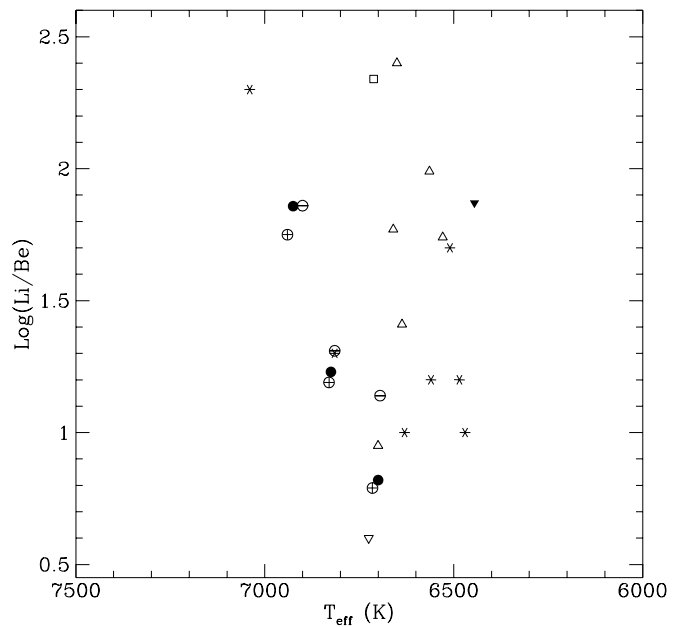,height=9cm}
}
\caption{Logarithm of the lithium to beryllium ratio as a function of
the effective temperature. The predictions are given with the same symbols 
as in Fig.~\ref{livsteff}. The asterisks and inverted triangles
correspond to the observations and upper limits in the Hyades 
(Boesgaard \& Budge 1989), and the square and triangles to observations 
and upper limits in field stars with [Fe/H] between +0.09 and -0.39 
(Stephens et al. 1997)
\label{be}
}
\end{figure}

\noindent {\bf {Stars cooler than $\sim$ 6400 K}}
\vspace{0.1cm}

\noindent The maximum of the Li abundance at $\sim$ 6400 K indicates
that the ``extra'' process reaches there its full efficiency, which must 
be sufficient to lead to solid body rotation in the Sun (at the solar age).
We stress once more that a self-consistent calculation of Li destruction 
requires a description of that mechanism.
It is however possible to estimate the lower limit of Li destruction
by calculating
the meridional circulation present in solid body rotators (filled 
triangles in Figs.~6, 7 and 8); 
this must be viewed only as an upper limit on the Li abundance for stars of 
this domain.
\vspace{0.1cm}

Two important remarks have to be done on the transition region. 

Firstly, it was shown by Balachandran (1995) 
that the location of the Li dip in different clusters depends
on the effective temperature {\it on the ZAMS} and is thus
independent of metallicity. Our explanation is entirely consistent
with this observation\footnote{Let us note that this is also the case for 
several of the mechanisms which have been suggested so far, for the
reason mentioned here.} since for low mass stars, the size of the
external convection zone is directly related to the effective temperature.
It is then the former which controls magnetic activity (and thus,
the spin-down) as well as the onset of the ``efficient'' mechanism
for angular momentum transport.

Secondly, the measures of Li abundances in the Hyades stars
of 6600$\ga T_{\rm eff}({\rm K}) \ga$6200 exhibit a large dispersion while it is 
much smaller in cooler stars.
This can be again linked to different rotational histories of stars of
the same mass.  Indeed, when these are young, the fast rotators have a larger 
convective zone than their slower counterparts (see Table 1 and 
Fig.~\ref{basezc} top). 
Therefore, in the early stages meridional circulation could be lower than 
expected, leading to a smaller Li destruction. 
Even though the magnitude of this change of $m_{zc}$ may seem
small, it is not negligible compared to the size of the transition
region.
A detailed description of angular momentum transport
by the ``efficient'' process is however required in order to quantify the
magnitude of this effect. 
\section{Conclusions}
Assuming rapid rotation and
using a self-consistent description for the transport of angular
momentum and of chemicals by meridional circulation and shear
instabilities (cf. Zahn 1992, Talon \& Zahn 1997), Talon et al. (1997)
successfully explained the C and N anomalies observed in some
B stars.

At the same time, it was shown (Matias \& Zahn 1997) that this
description applied to the transport of angular momentum in the Sun
is incomplete, leading to large $\Omega$ gradients which
are not observed. Another transport mechanism must thus be
invoked in low mass stars. 

At this point, 2 questions remain :
firstly, the nature of that transport mechanism
has to be determined unambiguously and secondly, the 
location of the transition
between the regime which is relevant for massive stars and the one
which is relevant for low mass stars has to be identified.

In this paper, we addressed that second question.
We presented numerical calculations of Li destruction
due to rotational mixing using the {\it same} description as Talon et al. 
used for more massive stars and the same free parameters.
We showed that this clearly reproduces the hot side of the Li dip.
Let us recall that the destruction of lithium is then
due solely to rotational mixing enhanced by the
spin down of the outer layers. Stars hotter than 7000 K also
undergo rotational mixing, but it is much milder due to the weak
differential rotation.

The rise of Li abundances on the right side of the dip is not explained
within this framework.
We propose that it is linked to the appearance of another transport 
mechanism for angular momentum which reduces the magnitude of the
meridional circulation and shears, leading to the observed
diminution of Li destruction on the red side of the Li dip.
This mechanism is known to occur in the Sun where it is responsible for the 
flat rotation profile. 
\begin{acknowledgements}
We would like to thank Jean-Paul Zahn for his careful reading of
this manuscript.
This work was supported by grants from the GDR 131 ``Structure Interne des
Etoiles et Plan\`etes Geantes'' (CNRS).
S.T. gratefully acknowledges support from NSERC of Canada and from
the French Minist\`ere des Affaires \'Etrang\`eres.
\end{acknowledgements}
\appendix
\section{Input physics}

The calculations are performed with the Toulouse-Geneva stellar evolution 
code, in which the equations described in \S 3.1 have been 
implemented.
Transport processes are engaged 
on the arrival on the zero age main sequence. 
The stellar models were obtained with the following input micro-physics :

\begin{itemize}
\item Microscopic diffusion : The tables of collision integrals by
Paquette et al. (1986) are used to calculate the diffusion coefficients.
\item Opacities : The radiative opacities for the interior are taken 
from Iglesias \& Rogers (1996). The low-temperature opacities 
are from Alexander \& Ferguson (1994) which account for a wide variety of 
atomic and molecular species. 
\item Nuclear cross-sections :  All the thermonuclear reaction rates
are due to Caughlan \& Fowler (1988), with the exception of the
$^{17}$O(p,$\gamma)^{18}$F and $^{17}$O(p,$\alpha)^{14}$N for which we
adopt the values from Landr\'e et al. (1990). Screening factors for the
reaction rates are taken into account according to the analytical 
prescription by Graboske et al. (1973). 
\item Abundances : 
We use the proto-solar lithium abundance derived from 
carbonaceous chondrites by Anders \& Grevesse (1989), 
log N(Li)=3.31$\pm$0.04, as cosmic value.
This Li abundance is consistent with the photospheric Li abundances in 
F stars of young open clusters like $\alpha$Per 
(Balachandran et al. 1996) 
and the Pleiades (Soderblom et al. 1993). 

We use [Fe/H]=+0.12 (Cayrel de Strobel et al. 1997)
for the logarithm of the number abundances of iron to hydrogen relative to 
the solar values.  
The initial helium content is determined by 
Y$=0.227 + (\Delta$Y$/ \Delta$Z)Z, 
where the value for $\Delta$Y$/ \Delta$Z we use (2.524) was obtained by
calibration of the solar models with the micro-physics described above and 
a solar metal abundance Z/X=0.0244 (Grevesse \& Noels 1993).
The relative ratios for the heavy elements correspond to the mixture by 
Grevesse \& Noels (1993). 
We take the same isotopic ratios than Maeder (1983).
\item Convection : Turbulent  convection  is  described  by the classical
mixing-length theory of B\"ohm-Vitense (1958). Calibration of the solar
model gave us a value of 1.6 for the free parameter $\alpha $, ratio
of  the  mixing-length  to  the  pressure  scale  height. 
Neither overshooting, nor convective penetration have been considered in this
work.
\end{itemize}

\end{document}